\documentclass[a4paper,12pt]{article}

\usepackage{amsmath}
\usepackage{amssymb}
\usepackage{amsthm}

\newtheorem{theorem}{Theorem}[section]
\newtheorem{lemma}[theorem]{Lemma}
\newtheorem{coro}[theorem]{Corollary}
\newtheorem{proposition}[theorem]{Proposition}

\theoremstyle{definition}
\newtheorem{definition}[theorem]{Definition}

\newtheorem{example}[theorem]{Example}
\newtheorem{remark}[theorem]{Remark}
\newlength{\fuyasu}
\def\cd{\cdots}
\def\ot{\otimes}

\def\bp{{\bf p}}
\def\by{{\bf y}}
\def\P{{\mathcal P}}

\def\veps{\varepsilon}
\def\vphi{\varphi}

\def\Z{{\mathbb Z}}
\def\Zn{\Z_{\ge0}}
\def\etd{\tilde{e}}
\def\ftd{\tilde{f}}

\def\ft#1{\tilde{f}_{#1}}
\newcommand{\dominocrossbox}[3]{%
  \multiput(0,0)(1,0){2}{\line(0,1){2}}
  \multiput(0,0)(0,1){3}{\line(1,0){1}}
  \put(0,0){\makebox(1,1){${#2}$}}
  \put(0,1){\makebox(1,1){${#1}$}}
  \put(1,0.5){\makebox(1,1){${\otimes}$}}
  \multiput(2,0.5)(1,0){2}{\line(0,1){1}}
  \multiput(2,0.5)(0,1){2}{\line(1,0){1}}
  \put(2,0.5){\makebox(1,1){${#3}$}}
}
\newcommand{\boxcrossdomino}[3]{%
  \multiput(2,0)(1,0){2}{\line(0,1){2}}
  \multiput(2,0)(0,1){3}{\line(1,0){1}}
  \put(2,0){\makebox(1,1){${#3}$}}
  \put(2,1){\makebox(1,1){${#2}$}}
  \put(1,0.5){\makebox(1,1){${\otimes}$}}
  \multiput(0,0.5)(1,0){2}{\line(0,1){1}}
  \multiput(0,0.5)(0,1){2}{\line(1,0){1}}
  \put(0,0.5){\makebox(1,1){${#1}$}}
}
\newcommand{\baronetwolast}[4]{%
\put(1,0){\line(1,0){5}}
\put(1,1){\line(1,0){5}}
\put(1,0){\line(0,1){1}}
\put(2,0){\line(0,1){1}}
\put(3,0){\line(0,1){1}}
\put(5,0){\line(0,1){1}}
\put(6,0){\line(0,1){1}}
\put(1,0){\makebox(1,1){$#1$}}
\put(2,0){\makebox(1,1){$#2$}}
\put(3,0){\makebox(2,1){$\cd$}}
\put(5,0){\makebox(1,1){$#3$}}
  \put(6,0){\makebox(1,1){${\otimes}$}}
  \multiput(7,0)(1,0){2}{\line(0,1){1}}
  \multiput(7,0)(0,1){2}{\line(1,0){1}}
  \put(7,0){\makebox(1,1){${#4}$}}
}
\newcommand{\barmiddle}[4]{%
  \multiput(0,0)(1,0){2}{\line(0,1){1}}
  \multiput(0,0)(0,1){2}{\line(1,0){1}}
  \put(0,0){\makebox(1,1){${#1}$}}
  \put(1,0){\makebox(1,1){${\otimes}$}}
\put(2,0){\line(1,0){7}}
\put(2,1){\line(1,0){7}}
\multiput(2,0)(1,0){2}{\line(0,1){1}}
\put(5,0){\line(0,1){1}}
\put(6,0){\line(0,1){1}}
\multiput(8,0)(1,0){2}{\line(0,1){1}}
\put(2,0){\makebox(1,1){$\cd$}}
\put(3,0){\makebox(2,1){$#2$}}
\put(5,0){\makebox(1,1){$#3$}}
\put(6,0){\makebox(2,1){$#4$}}
\put(8,0){\makebox(1,1){$\cd$}}
}
\newcommand{\boxbaronetwolast}[4]{%
\put(3,0){\line(1,0){7}}
\put(3,1){\line(1,0){7}}
\put(3,0){\line(0,1){1}}
\put(4,0){\line(0,1){1}}
\put(5,0){\line(0,1){1}}
\put(8,0){\line(0,1){1}}
\put(10,0){\line(0,1){1}}
\put(3,0){\makebox(1,1){$#2$}}
\put(4,0){\makebox(1,1){$#3$}}
\put(5,0){\makebox(3,1){$\cd$}}
\put(8,0){\makebox(2,1){$#4$}}
  \put(2,0){\makebox(1,1){${\otimes}$}}
  \multiput(1,0)(1,0){2}{\line(0,1){1}}
  \multiput(1,0)(0,1){2}{\line(1,0){1}}
  \put(1,0){\makebox(1,1){${#1}$}}
}
\newcommand{\dominocrossbar}[5]{%
  \multiput(0,0)(1,0){2}{\line(0,1){2}}
  \multiput(0,0)(0,1){3}{\line(1,0){1}}
  \put(0,0){\makebox(1,1){${#2}$}}
  \put(0,1){\makebox(1,1){${#1}$}}
  \put(1,0.5){\makebox(1,1){${\otimes}$}}
\put(2,0.5){\line(1,0){13}}
\put(2,1.5){\line(1,0){13}}
\put(2,0.5){\line(0,1){1}}
\put(3,0.5){\line(0,1){1}}
\put(4,0.5){\line(0,1){1}}
\put(13,0.5){\line(0,1){1}}
\put(15,0.5){\line(0,1){1}}
\put(2,0.5){\makebox(1,1){$#3$}}
\put(3,0.5){\makebox(1,1){$#4$}}
\put(8,0.5){\makebox(1,1){$\cd$}}
\put(13,0.5){\makebox(2,1){$#5$}}
}
\newcommand{\dominocrossbarx}[8]{%
  \multiput(0,0)(1,0){2}{\line(0,1){2}}
  \multiput(0,0)(0,1){3}{\line(1,0){1}}
  \put(0,0){\makebox(1,1){${#2}$}}
  \put(0,1){\makebox(1,1){${#1}$}}
  \put(1,0.5){\makebox(1,1){${\otimes}$}}
\put(2,0.5){\line(1,0){13}}
\put(2,1.5){\line(1,0){13}}
\put(2,0.5){\line(0,1){1}}
\put(3,0.5){\line(0,1){1}}
\put(4,0.5){\line(0,1){1}}
\put(6,0.5){\line(0,1){1}}
\put(8,0.5){\line(0,1){1}}
\put(9,0.5){\line(0,1){1}}
\put(11,0.5){\line(0,1){1}}
\put(13,0.5){\line(0,1){1}}
\put(15,0.5){\line(0,1){1}}
\put(2,0.5){\makebox(1,1){$#3$}}
\put(3,0.5){\makebox(1,1){$#4$}}
\put(4,0.5){\makebox(2,1){$\cd$}}
\put(6,0.5){\makebox(2,1){$#5$}}
\put(8,0.5){\makebox(1,1){$#6$}}
\put(9,0.5){\makebox(2,1){$#7$}}
\put(11,0.5){\makebox(2,1){$\cd$}}
\put(13,0.5){\makebox(2,1){$#8$}}
}
\newcommand{\dominocrossbarxx}[7]{%
  \multiput(0,0)(1,0){2}{\line(0,1){2}}
  \multiput(0,0)(0,1){3}{\line(1,0){1}}
  \put(0,0){\makebox(1,1){${#2}$}}
  \put(0,1){\makebox(1,1){${#1}$}}
  \put(1,0.5){\makebox(1,1){${\otimes}$}}
\put(2,0.5){\line(1,0){13}}
\put(2,1.5){\line(1,0){13}}
\put(2,0.5){\line(0,1){1}}
\put(3,0.5){\line(0,1){1}}
\put(5,0.5){\line(0,1){1}}
\put(7,0.5){\line(0,1){1}}
\put(8,0.5){\line(0,1){1}}
\put(10,0.5){\line(0,1){1}}
\put(13,0.5){\line(0,1){1}}
\put(15,0.5){\line(0,1){1}}
\put(2,0.5){\makebox(1,1){$#3$}}
\put(3,0.5){\makebox(2,1){$\cd$}}
\put(5,0.5){\makebox(2,1){$#4$}}
\put(7,0.5){\makebox(1,1){$#5$}}
\put(8,0.5){\makebox(2,1){$#6$}}
\put(11,0.5){\makebox(1,1){$\cd$}}
\put(13,0.5){\makebox(2,1){$#7$}}
}
\newcommand{\dominocrossbarxxx}[8]{%
  \multiput(0,0)(1,0){2}{\line(0,1){2}}
  \multiput(0,0)(0,1){3}{\line(1,0){1}}
  \put(0,0){\makebox(1,1){${#2}$}}
  \put(0,1){\makebox(1,1){${#1}$}}
  \put(1,0.5){\makebox(1,1){${\otimes}$}}
\put(2,0.5){\line(1,0){13}}
\put(2,1.5){\line(1,0){13}}
\put(2,0.5){\line(0,1){1}}
\put(3,0.5){\line(0,1){1}}
\put(5,0.5){\line(0,1){1}}
\put(6,0.5){\line(0,1){1}}
\put(8,0.5){\line(0,1){1}}
\put(9,0.5){\line(0,1){1}}
\put(11,0.5){\line(0,1){1}}
\put(12,0.5){\line(0,1){1}}
\put(14,0.5){\line(0,1){1}}
\put(15,0.5){\line(0,1){1}}
\put(2,0.5){\makebox(1,1){$\cd$}}
\put(3,0.5){\makebox(2,1){$#3$}}
\put(5,0.5){\makebox(1,1){$#4$}}
\put(6,0.5){\makebox(2,1){$#5$}}
\put(8,0.5){\makebox(1,1){$\cd$}}
\put(9,0.5){\makebox(2,1){$#6$}}
\put(11,0.5){\makebox(1,1){$#7$}}
\put(12,0.5){\makebox(2,1){$#8$}}
\put(14,0.5){\makebox(1,1){$\cd$}}
}
\newcommand{\barcrossdomino}[5]{%
\put(1,0.5){\line(1,0){5}}
\put(1,1.5){\line(1,0){5}}
\put(1,0.5){\line(0,1){1}}
\put(2,0.5){\line(0,1){1}}
\put(3,0.5){\line(0,1){1}}
\put(5,0.5){\line(0,1){1}}
\put(6,0.5){\line(0,1){1}}
\put(1,0.5){\makebox(1,1){$#1$}}
\put(2,0.5){\makebox(1,1){$#2$}}
\put(3,0.5){\makebox(2,1){$\cd$}}
\put(5,0.5){\makebox(1,1){$#3$}}
\put(6,0.5){\makebox(1,1){${\otimes}$}}
\multiput(7,0)(1,0){2}{\line(0,1){2}}
\multiput(7,0)(0,1){3}{\line(1,0){1}}
\put(7,0){\makebox(1,1){${#5}$}}
\put(7,1){\makebox(1,1){${#4}$}}
}
\newcommand{\singlebox}[1]{%
  \setlength{\unitlength}{5mm}
  \begin{picture}(1,1)(0,0.3)
  \multiput(0,0)(1,0){2}{\line(0,1){1}}
  \multiput(0,0)(0,1){2}{\line(1,0){1}}
  \put(0,0){\makebox(1,1){${#1}$}}
  \end{picture}
}
\newcommand{\domino}[2]{%
  \setlength{\unitlength}{5mm}
  \begin{picture}(1,2)(0,0.8)
  \multiput(0,0)(1,0){2}{\line(0,1){2}}
  \multiput(0,0)(0,1){3}{\line(1,0){1}}
  \put(0,0){\makebox(1,1){${#2}$}}
  \put(0,1){\makebox(1,1){${#1}$}}
  \end{picture}
}
\newcommand{\threeboxes}[3]{%
  \setlength{\unitlength}{5mm}
  \begin{picture}(3,1)(0,0.3)
  \multiput(0,0)(1,0){4}{\line(0,1){1}}
  \multiput(0,0)(0,1){2}{\line(1,0){3}}
  \put(0,0){\makebox(1,1){${#1}$}}
  \put(1,0){\makebox(1,1){${#2}$}}
  \put(2,0){\makebox(1,1){${#3}$}}
  \end{picture}
}
\newcommand{\sixlettersinbar}[6]{%
  \setlength{\unitlength}{5mm}
  \begin{picture}(13,1)(2,0.3)
\put(2,0){\line(1,0){13}}
\put(2,1){\line(1,0){13}}
\put(2,0){\line(0,1){1}}
\put(3,0){\line(0,1){1}}
\put(5,0){\line(0,1){1}}
\put(6,0){\line(0,1){1}}
\put(8,0){\line(0,1){1}}
\put(9,0){\line(0,1){1}}
\put(11,0){\line(0,1){1}}
\put(12,0){\line(0,1){1}}
\put(14,0){\line(0,1){1}}
\put(15,0){\line(0,1){1}}
\put(2,0){\makebox(1,1){$\cd$}}
\put(3,0){\makebox(2,1){$#1$}}
\put(5,0){\makebox(1,1){$#2$}}
\put(6,0){\makebox(2,1){$#3$}}
\put(8,0){\makebox(1,1){$\cd$}}
\put(9,0){\makebox(2,1){$#4$}}
\put(11,0){\makebox(1,1){$#5$}}
\put(12,0){\makebox(2,1){$#6$}}
\put(14,0){\makebox(1,1){$\cd$}}
  \end{picture}
}
\newcommand{\fivelettersinbar}[5]{%
  \setlength{\unitlength}{5mm}
  \begin{picture}(13,1)(2,0.3)
\put(2,0){\line(1,0){13}}
\put(2,1){\line(1,0){13}}
\put(2,0){\line(0,1){1}}
\put(3,0){\line(0,1){1}}
\put(6,0){\line(0,1){1}}
\put(8,0){\line(0,1){1}}
\put(9,0){\line(0,1){1}}
\put(11,0){\line(0,1){1}}
\put(14,0){\line(0,1){1}}
\put(15,0){\line(0,1){1}}
\put(2,0){\makebox(1,1){$#1$}}
\put(3.5,0){\makebox(2,1){$\cd$}}
\put(6,0){\makebox(2,1){$#2$}}
\put(8,0){\makebox(1,1){$#3$}}
\put(9,0){\makebox(2,1){$#4$}}
\put(12,0){\makebox(1,1){$\cd$}}
\put(14,0){\makebox(1,1){$#5$}}
  \end{picture}
}
\newcommand{\threelettersinbar}[3]{%
  \setlength{\unitlength}{5mm}
  \begin{picture}(13,1)(2,0.3)
\put(2,0){\line(1,0){13}}
\put(2,1){\line(1,0){13}}
\put(2,0){\line(0,1){1}}
\put(3,0){\line(0,1){1}}
\put(13,0){\line(0,1){1}}
\put(14,0){\line(0,1){1}}
\put(15,0){\line(0,1){1}}
\put(2,0){\makebox(1,1){$#1$}}
\put(8,0){\makebox(1,1){$\cd$}}
\put(13,0){\makebox(1,1){$#2$}}
\put(14,0){\makebox(1,1){$#3$}}
  \end{picture}
}
\newcommand{\sixlettersinbarx}[6]{%
  \setlength{\unitlength}{5mm}
  \begin{picture}(13,1)(2,0.3)
\put(2,0){\line(1,0){13}}
\put(2,1){\line(1,0){13}}
\put(2,0){\line(0,1){1}}
\put(3,0){\line(0,1){1}}
\put(5,0){\line(0,1){1}}
\put(7,0){\line(0,1){1}}
\put(8,0){\line(0,1){1}}
\put(10,0){\line(0,1){1}}
\put(13,0){\line(0,1){1}}
\put(14,0){\line(0,1){1}}
\put(15,0){\line(0,1){1}}
\put(2,0){\makebox(1,1){$#1$}}
\put(3,0){\makebox(2,1){$\cd$}}
\put(5,0){\makebox(2,1){$#2$}}
\put(7,0){\makebox(1,1){$#3$}}
\put(8,0){\makebox(2,1){$#4$}}
\put(11,0){\makebox(1,1){$\cd$}}
\put(13,0){\makebox(1,1){$#5$}}
\put(14,0){\makebox(1,1){$#6$}}
  \end{picture}
}
\newcommand{\threelettersinbarx}[3]{%
  \setlength{\unitlength}{5mm}
  \begin{picture}(13,1)(2,0.3)
\put(2,0){\line(1,0){13}}
\put(2,1){\line(1,0){13}}
\put(2,0){\line(0,1){1}}
\put(3,0){\line(0,1){1}}
\put(4,0){\line(0,1){1}}
\put(14,0){\line(0,1){1}}
\put(15,0){\line(0,1){1}}
\put(2,0){\makebox(1,1){$#1$}}
\put(3,0){\makebox(1,1){$#2$}}
\put(9,0){\makebox(1,1){$\cd$}}
\put(3,0){\makebox(1,1){$#2$}}
\put(14,0){\makebox(1,1){$#3$}}
  \end{picture}
}

\newcommand{\carrierpicture}[7]{%
  \setlength{\unitlength}{3mm}
  \begin{picture}(4,4)(0,0)
  \multiput(0,1)(1,0){2}{\line(0,1){2}}
  \multiput(3,1)(1,0){2}{\line(0,1){2}}
  \multiput(1.5,-0.5)(1,0){2}{\line(0,1){1}}
  \multiput(1.5,3.5)(1,0){2}{\line(0,1){1}}
  \multiput(0,1)(3,0){2}{\line(1,0){1}}
  \multiput(1.5,-0.5)(0,4){2}{\line(1,0){1}}
  \put(1.1,2){\vector(1,0){1.8}}
  \put(2,3.4){\vector(0,-1){2.8}}
  \put(0,2){\makebox(1,1){${\scriptstyle {#1}}$}}
  \put(0,1){\makebox(1,1){${\scriptstyle {#2}}$}}
  \put(1.5,3.5){\makebox(1,1){${\scriptstyle {#3}}$}}
  \put(1.5,-0.5){\makebox(1,1){${\scriptstyle {#4}}$}}
  \put(3,2){\makebox(1,1){${\scriptstyle {#5}}$}}
  \put(3,1){\makebox(1,1){${\scriptstyle {#6}}$}}
  \put(1,-1.7){\makebox(2,1){${\scriptstyle {#7}}$}}
  \put(0,0.2){\makebox(1,1){$\circ$}}
  \put(3,0.2){\makebox(1,1){$\circ$}}
  \end{picture}
}
\newcommand{\carrierpicturedouble}[9]{%
  \setlength{\unitlength}{3mm}
  \begin{picture}(7,4)(0,1.5)
  \multiput(0,2)(3,0){3}{\line(0,1){2}}
  \multiput(1,2)(3,0){3}{\line(0,1){2}}
  \multiput(0,2)(3,0){3}{\line(1,0){1}}
  \multiput(1.5,0.5)(3,0){2}{\line(0,1){1}}
  \multiput(2.5,0.5)(3,0){2}{\line(0,1){1}}
  \multiput(1.5,0.5)(3,0){2}{\line(1,0){1}}
  \multiput(1.5,4.5)(3,0){2}{\line(0,1){1}}
  \multiput(2.5,4.5)(3,0){2}{\line(0,1){1}}
  \multiput(1.5,4.5)(3,0){2}{\line(1,0){1}}
  \multiput(1.1,3)(3,0){2}{\vector(1,0){1.8}}
  \multiput(2,4.4)(3,0){2}{\vector(0,-1){2.8}}
  \multiput(0,1.2)(3,0){3}{\makebox(1,1){$\circ$}}
  \put(0,3){\makebox(1,1){${\scriptstyle {#1}}$}}
  \put(0,2){\makebox(1,1){${\scriptstyle {#2}}$}}
  \put(1.5,4.5){\makebox(1,1){${\scriptstyle {#3}}$}}
  \put(1.5,0.5){\makebox(1,1){${\scriptstyle {#4}}$}}
  \put(3,3){\makebox(1,1){${\scriptstyle {#5}}$}}
  \put(3,2){\makebox(1,1){${\scriptstyle {#6}}$}}
  \put(4.5,0.5){\makebox(1,1){${\scriptstyle {#7}}$}}
  \put(6,2){\makebox(1,1){${\scriptstyle {#8}}$}}
  \put(2.5,-1.7){\makebox(2,1){${\scriptstyle {#9}}$}}
\end{picture}
}
\newcommand{\carrierpictureseq}{%
  \setlength{\unitlength}{3mm}
  \begin{picture}(22,6)(0,0)
  \multiput(0,2)(3,0){8}{\line(0,1){2}}
  \multiput(1,2)(3,0){8}{\line(0,1){2}}
  \multiput(0,2)(3,0){8}{\line(1,0){1}}
  \multiput(1.5,0.5)(3,0){7}{\line(0,1){1}}
  \multiput(2.5,0.5)(3,0){7}{\line(0,1){1}}
  \multiput(1.5,0.5)(3,0){7}{\line(1,0){1}}
  \multiput(1.5,4.5)(3,0){7}{\line(0,1){1}}
  \multiput(2.5,4.5)(3,0){7}{\line(0,1){1}}
  \multiput(1.5,4.5)(3,0){7}{\line(1,0){1}}
  \multiput(1.1,3)(3,0){7}{\vector(1,0){1.8}}
  \multiput(2,4.4)(3,0){7}{\vector(0,-1){2.8}}
  \put(1.5,4.5){\makebox(1,1){${\scriptstyle 5}$}}
  \put(4.5,4.5){\makebox(1,1){${\scriptstyle 5}$}}
  \put(7.5,4.5){\makebox(1,1){${\scriptstyle 4}$}}
  \put(10.5,4.5){\makebox(1,1){${\scriptstyle 3}$}}
  \put(13.5,4.5){\makebox(1,1){${\scriptstyle 2}$}}
  \put(4.5,0.5){\makebox(1,1){${\scriptstyle 5}$}}
  \put(7.5,0.5){\makebox(1,1){${\scriptstyle 5}$}}
  \put(10.5,0.5){\makebox(1,1){${\scriptstyle 4}$}}
  \put(13.5,0.5){\makebox(1,1){${\scriptstyle 2}$}}
  \put(16.5,0.5){\makebox(1,1){${\scriptstyle 2}$}}
  \multiput(3,3)(3,0){5}{\makebox(1,1){${\scriptstyle 2}$}}
  \put(0,2){\makebox(1,1){${\scriptstyle 2}$}}
  \multiput(3,2)(3,0){2}{\makebox(1,1){${\scriptstyle 5}$}}
  \put(9,2){\makebox(1,1){${\scriptstyle 4}$}}
  \multiput(12,2)(3,0){4}{\makebox(1,1){${\scriptstyle 3}$}}
  \multiput(0,1.2)(3,0){8}{\makebox(1,1){$\circ$}}
  \end{picture}
}

\title{Separation of colour degree of freedom
from dynamics in a soliton cellular automaton}
\author{Taichiro Takagi\\
\normalsize
\em Department of Applied Physics, National Defense Academy,\\
\normalsize
\em Kanagawa 239-8686, Japan}
\date{}
\begin{document}
\maketitle
\begin{abstract}
We present an algorithm to reduce the coloured box-ball system, 
a one dimensional integrable cellular automaton described by
motions of several colour (kind) of balls, into a 
simpler monochrome system.
This algorithm extracts the colour degree of freedom of the
automaton as a word which turns out
to be a conserved quantity of this dynamical system.
It is based on the theory of crystal basis and in particular
on the tensor products of $sl_n$ crystals of symmetric 
and anti-symmetric tensor representations.
\end{abstract}
\section{Introduction}\label{sec:1}
The soliton cellular automaton by Takahashi and Satsuma \cite{TS} is
a discrete dynamical system
related to the KdV equation.
This automaton can be described by motions of finite number of
balls on an array of boxes, so it would have been called
a {\em box and ball system} or a {\em box-ball system} for short.
The original box-ball system in \cite{TS} has only one kind of balls.
Generalizations to the systems with several kinds (colours) of balls
were introduced and studied \cite{Th,TNS,TTM}.
We shall call these systems {\em coloured systems}, and
the original one the (basic) {\em monochrome system}.
Some years ago, 
a connection between these coloured systems and
crystal basis theory \cite{K1,K2}
was found \cite{FOY,HHIKTT}.
By the algebraic (combinatorial) methods in the 
crystal basis theory,
the scattering rules of solitons in the coloured
systems \cite{TNS} were explained.

In contrast to this success in the study on the scattering rules,
there remain some problems on the construction of
their conserved quantities, or
of their general $N$-soliton solutions.
There are studies
on this subject for
the coloured systems by
analytical methods \cite{TNS,TTM} or by
combinatorial methods \cite{FOY,F}.
However the expressions for
conserved quantities (or $N$-soliton solutions)
in the former are rather complicated, and
in both cases we did not know whether
we have already obtained the list of all conserved 
quantities.
It is clear that the difficulties of finding a 
simple and complete description of
all conserved quantities are due to the colour degree of freedom of
these systems.
In fact for the monochrome system the
construction of all conserved quantities 
and the linearization of its dynamics
{(which are equivalent to the construction of
general $N$-soliton solutions)}
have been done \cite{TTS,Tg}
(but in the {\em basic} case, explained below).

In this paper we present an algorithm to separate this
colour degree of freedom from dynamics in the coloured systems.
The idea used here is
an isomorphism between 
tensor products of crystals of symmetric and anti-symmetric
tensor representations \cite{HKOTY}.
Our algorithm reduces a coloured system
to a monochrome system, giving a {\em word} 
(finite sequence of colours of the balls) at the same time.
We shall show that
this word itself is a conserved quantity, i.e.~it does not
change under the time evolution of the automaton.

Throughout this paper we shall call
the automata with all the cells (boxes) having capacity one
{\em basic} systems \cite{Th,TNS,FOY}, 
and those with cells of various capacities
{\em inhomogeneous} systems \cite{TTM,HHIKTT,F}.
Our algorithm of separation of colour degree of freedom works
in both cases.
Clearly the former is a special case of the latter, so it is enough
to give a proof for the latter case only.
However we shall give a complete description of the proof for
the basic case 
first and then generalize it to the inhomogeneous case,
because we think it is more accessible for many readers.

The layout of this paper is as follows.
In section \ref{sec:2} we introduce 
the {basic} 
coloured system and explain how the algorithm of separation of
colour degree of freedom is conducted.
We use a description by a carrier of balls
which we call a decoding carrier.
In section \ref{sec:3} we recall basic notions in the theory of
$sl_n$ crystals and their tensor product decomposition by the
Littlewood-Richardson rule.
In section \ref{sec:4} we show explicit formulas for
the isomorphism between tensor products of crystals, and explain
the symmetric group generated by them.
Here the decoding carrier
is identified with an element of a crystal for the anti-symmetric
representation.
In section \ref{sec:5} we recall a description of coloured systems
by means of the crystals, and give a proof for the separation of colour
degree of freedom with the tools prepared in the previous sections.
In section \ref{sec:6} we generalize the proof to that for the 
inhomogeneous system.
Some examples and details of calculations are given in the
Appendices.

\section{The coloured system and the decoding carrier}\label{sec:2}

We recall the algorithm for the time evolution of
the basic coloured system \cite{Th, TNS}.
To adjust to the notations in crystal basis theory and
Young tableaux, we
call ``(a box containing) a ball with index $i$" simply
``(letter) $i (\geq 2)$", and identify ``an empty box" with
``(letter) $1$".
Fix an integer $n \geq 2$.
At time $t$ we have an infinite sequence of letters
$1,2,\ldots,n$.
The numbers of $2,\ldots,n$ are finite.
We denote this state by $\bp$, and write
the state at $t+1$ as $T(\bp)$, which means that 
we define a time evolution operator $T$ that
applies on $\bp$.
\begin{definition}[\cite{Th,TNS}]\label{def:oct22_1}
The time evolution operator $T$ is given by
$T=K_2 \cd K_n$ where $K_i$ are the operators
which work as:
\begin{enumerate}
\item Move every letter $i$ only once.
\item Exchange the leftmost $i$ with its nearest right $1$.
\item Exchange the leftmost $i$ 
among the rest of the $i$'s
with its nearest right $1$.
\item Repeat this procedure until all of the $i$'s are moved.
\end{enumerate}
\end{definition}
\begin{remark}
The numbering of balls is opposite to those in \cite{Th,TNS}.
\end{remark}
%
\begin{example}\label{ex:oct3_1}
Here is an example of the time evolution of the monochrome system.
\begin{verbatim}
       t=0   ....22222......222.2.............
       t=1   .........22222....2.222..........
       t=2   ..............2222.2...2222......
       t=3   ..................2.222....22222.
\end{verbatim}
We denoted letter $1$ (empty box) by $``."$.
\end{example}
\begin{example}\label{ex:oct3_2}
Here is an example of the time evolution of the coloured system
which has four kinds of colours ``2,3,4,5".
\begin{verbatim}
       t=0   55432.....542....2...............
       t=1   .....55432...542..2..............
       t=2   ..........55432.54.22............
       t=3   ...............5435..54222.......
\end{verbatim}
\end{example}
We introduce another operator $T_\natural$.
For this purpose we consider
a carrier of balls which we call a {\em decoding carrier}.
The capacity of this carrier is two, but
it can not be empty, so it always has at least one ball.
It can not have balls of the same kind at once.
It loads and/or unloads balls when it passes by each box.
The loading-unloading processes  are
depicted as follows.
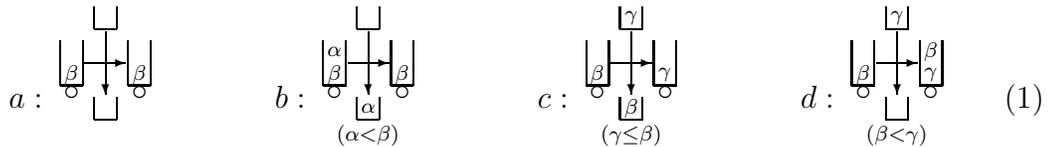
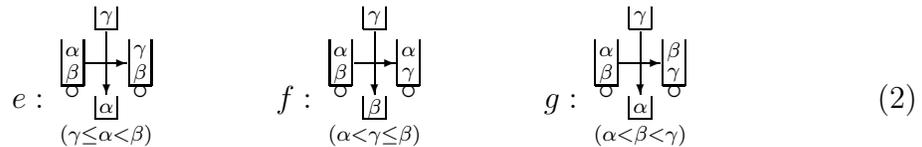
\begin{equation}\label{eq:oct10_11}
a: \;
\carrierpicture{}{\beta}{}{}{}{\beta}{} \qquad\qquad
b: \;
\carrierpicture{\alpha}{\beta}{}{\alpha}{}{\beta}{(\alpha < \beta)} 
\qquad\qquad
c: \;
\carrierpicture{}{\beta}{\gamma}{\beta}{}{\gamma}{(\gamma \leq \beta)} \qquad\qquad
d: \;
\carrierpicture{}{\beta}{\gamma}{}{\beta}{\gamma}{(\beta < \gamma)}
\end{equation}
\begin{equation}\label{eq:oct10_10}
e: \;
\carrierpicture{\alpha}{\beta}{\gamma}{\alpha}{\gamma}{\beta}{(\gamma \leq \alpha < \beta)} \qquad\qquad
f: \;
\carrierpicture{\alpha}{\beta}{\gamma}{\beta}{\alpha}{\gamma}{(\alpha < \gamma \leq \beta)} \qquad\qquad
g: \;
\carrierpicture{\alpha}{\beta}{\gamma}{\alpha}{\beta}{\gamma}{(\alpha < \beta < \gamma)}
\end{equation}
\vspace{5mm}
\par\noindent
For instance in the process $e$, a carrier containing (two balls
with indices) $\alpha$ and $\beta$ comes from the left to
a box containing $\gamma$.
It picks up the $\gamma$, puts the $\alpha$ into the box and
goes to the right.
Note that we are assuming $\alpha, \beta, \gamma \geq 2$ here,
and if we interpret a vacancy as letter ``1",
the processes $a$ and $b$ become special cases of $e$, and
$c$ and $d$ become those of $f$ and $g$ respectively.
We distinguished the processes $e$ and $g$ to adjust to
the notations used later, although they are formally the same.

Now we define the operator $T_\natural$.
Let the the decoding carrier always have
a ``2" at the beginning.
The carrier runs from left to right along the automaton state $\bp$, and
changes it to another state which we call $T_\natural(\bp)$.
It puts the ``2" into the automaton state and
takes off a letter $\geq 2$ from the state.
\begin{example}\label{ex:oct10_12}
In the following picture
the decoding carrier runs along the automaton state 
$55432..$, changes it into the state
$.55422.$ and takes off the letter 3.
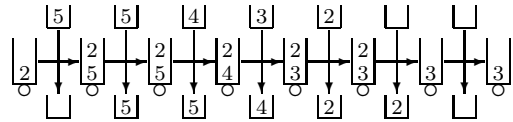
\begin{displaymath}
\carrierpictureseq 
\end{displaymath}
\end{example}
\begin{example}\label{ex:oct5_1}
The following {diagram} shows how an automaton state
(the first row in Example \ref{ex:oct3_2})
will be changed by applying $T_\natural$ repeatedly.
\begin{verbatim}
       s=0   55432.....542....2............... 2
       s=1   .55422.....532...4............... 4
       s=2   ..55222.....432..5............... 5
       s=3   ...52222....543...2.............. 2
       s=4   ....22222...554...3.............. 3
       s=5   ....22222....552..4.............. 4
       s=6   ....22222.....522.5.............. 5
       s=7   ....22222......2225.............. 5
       s=8   ....22222......222.2............. 
\end{verbatim}
Here $s$ is the number of times we have run the 
decoding carrier.
The number attached at the end of each row (except the last one)
shows the letter which will be taken off from that row.
Note that the automaton state in the
last row is equal to the first row in Example \ref{ex:oct3_1}.
\end{example}
Suppose, as in Example \ref{ex:oct5_1}, we could remove all the 
letters $\geq 3$ by applying $T_\natural$ finitely many times on $\bp$.
Let $\tilde{\bp}$ be the state obtained from $\bp$ after removing 
all the letters $\geq 3$.
We denote by $\by$ the word made by these 
removed letters in reverse order
($\by = 55432542$ for Example \ref{ex:oct5_1}).
We may write symbolically
\begin{equation}\label{eq:oct4_1}
\bp = \tilde{\bp} \oplus \by.
\end{equation}
In the following sections we shall show that for any state
$\bp$ we can decompose it into this form (\ref{eq:oct4_1}), and
that under the time evolution $T$
the following relation holds
\begin{equation}\label{eq:oct10_1}
T(\bp) = T(\tilde{\bp}) \oplus \by.
\end{equation}
We call this property of the coloured system a
{\em separation of colour degree of freedom from dynamics}.
For the above example we illustrate this property in Appendix \ref{app:a}.

Clearly this implies that
the word $\by$ here is a conserved quantity of the coloured system.
On the other hand the path $\tilde{\bp}$ 
is regarded as a state for the monochrome system, 
and it has its own conserved
quantities (and its dynamics can be linearized) \cite{Tg,TTS}.
{}From (\ref{eq:oct4_1}) and (\ref{eq:oct10_1}) they are
also conserved quantities of the original coloured system.
In this way we can construct all conserved quantities
(and can linearize the dynamics)
of the (basic) coloured system.
\section{Crystals and their tensor products}\label{sec:3}
The dynamics of the coloured system in the previous section can be
described by using $sl_n$ crystals.
In this section we {recall some properties of} the
$sl_n$ crystals in \cite{KN}.
The elements of the crystals $B_\lambda$ are given by
semi-standard Young tableaux 
with any shape $\lambda$ and with letters
$1,\cd, n$.
In particular we shall
consider the crystals $B_{(\ell)}$ and $B_{(1,1)}$
which we call $B_\ell$ and $B_\natural$ respectively.
They are given by
\begin{align}
B_\ell &= \left\{
\setlength{\unitlength}{5mm}
\begin{picture}(6,1.5)(0.5,0.3)
\put(1,0){\line(1,0){5}}
\put(1,1){\line(1,0){5}}
\put(1,0){\line(0,1){1}}
\put(2,0){\line(0,1){1}}
\put(3,0){\line(0,1){1}}
\put(5,0){\line(0,1){1}}
\put(6,0){\line(0,1){1}}
\put(1,0){\makebox(1,1){$\alpha_1$}}
\put(2,0){\makebox(1,1){$\alpha_2$}}
\put(3,0){\makebox(2,1){$\cd$}}
\put(5,0){\makebox(1,1){$\alpha_\ell$}}
\end{picture}
\Bigg|
1 \leq \alpha_1 \leq \alpha_2 \leq \cdots \leq \alpha_\ell \leq n
\right\},\label{eq:oct24_1}\\
B_\natural &= \left\{
\setlength{\unitlength}{5mm}
\begin{picture}(3,2)(-0.7,0.7)
\multiput(0,0)(1,0){2}{\line(0,1){2}}
\multiput(0,0)(0,1){3}{\line(1,0){1}}
\put(0,0){\makebox(1,1){${\beta}$}}
\put(0,1){\makebox(1,1){${\alpha}$}}
\end{picture}
\Bigg|
1 \leq \alpha < \beta \leq n
\right\},\nonumber
\end{align}
as sets.
We shall mainly use these special types of crystals in the 
description of the coloured systems.
\begin{remark}
They are special cases of the crystals for rectangular shape
Young tableaux.
Such crystals can also be regarded as
{\em perfect} 
crystals for affine Lie algebra $\widehat{sl}_n$ \cite{KMN,S}
and this fact may reduce some arguments below.
In this paper we would rather
avoid to use this difficult concept of perfect crystals.
\end{remark}

We recall basic notions in the theory of crystals.
See \cite{KN} for details.
Let $I=\{1,\cdots,n-1\}$ be the index set 
and $B$ an $sl_n$ crystal.
For any $i \in I$ there are maps
$\ftd_i$ from $B\sqcup \{0\}$ to $B \sqcup \{0\}$
and maps $\veps_i$ and $\vphi_i$ from $B$ to $\Zn$.
(We shall omit descriptions of the maps 
$\etd_i$ which are basically defined as $(\ftd_i)^{-1}$.)
It is always assumed that $\ftd_i 0 = 0$.
For $B =B_{\ell}$ or $B_\natural$
their definitions are given as follows.
\begin{enumerate}
\item
($B=B_\ell$)
If $b \in B$ has at least one $i$ then
$\ftd_i (b)$ is the one obtained by replacing 
the rightmost $i$ with $i+1$.
Otherwise, $\ftd_i (b)=0$. 
The maps $\vphi_i$ and $\veps_i$ are given by
$\vphi_i(b)=\# (i \mbox{'s in } b)$ and
$\veps_i(b)=\# (i+1 \mbox{'s in } b)$.
\item
($B=B_\natural$)
If $i$ appears in $b \in B$ and $i+1$ does not, then
$\ftd_i (b)$ is the one obtained by replacing 
$i$ with $i+1$. Otherwise, $\ftd_i (b)=0$. 
The maps $\vphi_i$ and $\veps_i$ are given by
\begin{align*}
\vphi_i(b)&= [\mbox{$i$ appears in $b$ and $i+1$ does not}],\\
\veps_i(b)&= [\mbox{$i+1$ appears in $b$ and $i$ does not}],
\end{align*}
where $[\mbox{true}]=1, \; [\mbox{false}]=0$.
\end{enumerate}
Also the functions $\ftd_i, \vphi_i, \veps_i$ 
for more general $B_\lambda$ can be
defined as in \cite{KN} but we do not use their explicit forms.
We simply note that there is a unique special element called the
{\em highest weight element}
in any $B_\lambda$, 
and every element of $B_\lambda$ can be
obtained by applying some sequence of $\ftd_i$'s on it.
The highest weight element in $B_\lambda$
is given by the Young tableau
of shape $\lambda$ with all the letters in its first row are
$1$'s, those in its second row are
$2$'s, and so on.

An important property of crystals is their tensor
product structure.
Let $\lambda_1$ and $\lambda_2$ be any two Young diagrams,
and $B_{\lambda_1}$ and $B_{\lambda_2}$ the associated $sl_n$ 
crystals. 
As a set the tensor product $B_{\lambda_1}\ot B_{\lambda_2}$
is simply a direct product.
The actions of the maps $\ft{i}$ 
on $B_{\lambda_1}\ot B_{\lambda_2}$ are given by
\begin{equation}\label{eq:oct28_2}
\ft{i}(b\ot b')=\left\{
\begin{array}{ll}
\ft{i} b\ot b'&\mbox{ if }\vphi_i(b) > \veps_i(b')\\
b\ot \ft{i} b'&\mbox{ if }\vphi_i(b)\le\veps_i(b').
\end{array}\right. 
\end{equation}
Here $0\ot b'$ and $b\ot 0$ should be understood as $0$.
The tensor product has a finite decomposition
\begin{equation}\label{eq:oct28_1}
B_{\lambda_1} \ot B_{\lambda_2} \simeq \bigoplus_\nu 
B_\nu^{\oplus m_{\nu}},
\end{equation}
which is described by the Littlewood-Richardson 
(LR) rule \cite{N}.
Here $\nu$'s are distinct Young diagrams and $m_{\nu} (\geq 1)$
is the multiplicity of $B_\nu$.
The $\simeq$ denotes the isomorphism of $sl_n$ crystals, 
i.e.~it commutes with the actions of $\ftd_i$.
Tensor products of three or more crystals are
defined again by using (\ref{eq:oct28_2}).
To define them such formulas as
$\veps_i(b \ot b') = \max(\veps_i(b),\veps_i(b)+\veps_i(b')-\vphi(b))$ and
$\vphi_i(b \ot b') = \max(\vphi_i(b'),\vphi_i(b)+\vphi_i(b')-\veps(b'))$
are also used.
The decomposition of such a tensor product is given by
applying (\ref{eq:oct28_1}) repeatedly.

\section{Crystal isomorphism and the symmetric group}
\label{sec:4}
Now let $\lambda_1$ and $\lambda_2$ be rectangular shape Young diagrams.
Then $m_{\nu}=1$ for any $\nu$ in (\ref{eq:oct28_1}) hence
we have a unique isomorphism of crystals 
$B_{\lambda_1} \otimes B_{\lambda_2}
\simeq B_{\lambda_2} \otimes B_{\lambda_1}$ \cite{SW,S} 
which is given by {\em tableaux products} \cite{Fl}.

If $\lambda_1 = \lambda_2$ the isomorphism is trivial, 
i.~e.~it is given by the identity map.
Below we give explicit forms for non-trivial isomorphism
between $B_\ell=B_{(\ell)}, B_1=B_{(1)}$ and $B_\natural = 
B_{(1,1)}$,
which will be used in the description of the basic coloured system.

Let $\iota: B_\ell \ot B_1 \stackrel{\sim}{\rightarrow} 
B_1 \ot B_\ell$ be the map for the $sl_n$ crystal isomorphism.
The decomposition by LR rule
is given by
$ B_\ell \ot B_1 \simeq B_{(\ell+1)} \oplus B_{(\ell,1)}$
and we have
\begin{displaymath}
\iota: 
\setlength{\unitlength}{5mm}
\begin{picture}(7,1)(1,0.3)
\baronetwolast{\alpha_1}{\alpha_2}{\alpha_\ell}{\beta}
\end{picture}
\stackrel{\sim}{\mapsto}
\begin{cases}
\begin{picture}(7,1)(1,0.3)
\boxbaronetwolast{\alpha_\ell}{\beta}{\alpha_1}{\alpha_{\ell-1}}
\end{picture} & \mbox{if $\beta \leq \alpha_1$, } \\
\begin{picture}(9,1)(0,0.3)
\barmiddle{\alpha_p}{\alpha_{p-1}}{\beta}{\alpha_{p+1}}
\end{picture} & \mbox{if $\beta > \alpha_1$,}
\end{cases}
\end{displaymath}
where $p$ is determined by the condition
$\alpha_p < \beta \leq \alpha_{p+1}$.
The upper (resp.~lower) one is associated with 
$B_{(\ell+1)}$ (resp.~$B_{(\ell,1)}$).

Let $\iota': B_\natural \ot B_1 \stackrel{\sim}{\rightarrow} 
B_1 \ot B_\natural$ be the map for the $sl_n$ crystal isomorphism.
The decomposition by LR rule is given by
$ B_\natural \ot B_1 \simeq B_{(2,1)} \oplus B_{(1,1,1)}$
and we have
\begin{displaymath}
\iota': 
\setlength{\unitlength}{5mm}
\begin{picture}(3,2)(0,0.8)
\dominocrossbox{\alpha}{\beta}{\gamma}
\end{picture}
\stackrel{\sim}{\mapsto}
\begin{cases}
\begin{picture}(3,2)(0,0.3)
\boxcrossdomino{\alpha}{\gamma}{\beta}
\end{picture} & \mbox{if $\gamma \leq \alpha,$} \\
\begin{picture}(3,2)(0,0.3)
\boxcrossdomino{\beta}{\alpha}{\gamma}
\end{picture} & \mbox{if $\alpha < \gamma \leq \beta,$} \\
\begin{picture}(3,2)(0,0.3)
\boxcrossdomino{\alpha}{\beta}{\gamma}
\end{picture} & \mbox{if $\beta < \gamma$.}
\end{cases}
\end{displaymath}
The first and the second ones are associated with
$B_{(2,1)}$, and the third one with $B_{(1,1,1)}$.

\begin{remark}
The three cases here correspond to the loading-unloading
processes of the decoding carrier which are {denoted} by
$e,f,g$ in (\ref{eq:oct10_10}).
Below we sometimes call an element of $B_\natural$
{\em a carrier} for this reason. 
\end{remark}

Let $\iota'': B_\ell \ot B_\natural \stackrel{\sim}{\rightarrow} 
B_\natural \ot B_\ell$ be map for the $sl_n$ crystal isomorphism.
The decomposition by LR rule is given by
$ B_\ell \ot B_\natural \simeq B_{(\ell+1,1)} \oplus B_{(\ell,1,1)}$
and we have
\begin{align*}
\iota'' & :
\setlength{\unitlength}{5mm}
\begin{picture}(7,2)(1,0.8)
\barcrossdomino{\alpha_1}{\alpha_2}{\alpha_\ell}{\beta}{\gamma}
\end{picture} 
\\
& \stackrel{\sim}{\mapsto}
\begin{cases}
\setlength{\unitlength}{5mm}
\begin{picture}(15,2)(0,0.8)
\dominocrossbarxxx{\alpha_i}{\alpha_j}
{\alpha_{i-1}}{\beta}{\alpha_{i+1}}
{\alpha_{j-1}}{\gamma}{\alpha_{j+1}}
\end{picture} & \mbox{if 
$\begin{array}{l}
\alpha_i < \beta \leq \alpha_{i+1}, \\
\alpha_j < \gamma \leq \alpha_{j+1},
\end{array}
$} \\
\setlength{\unitlength}{5mm}
\begin{picture}(15,2)(0,0.8)
\dominocrossbarxx{\alpha_i}{\beta}
{\alpha_{1}}{\alpha_{i-1}}{\gamma}{\alpha_{i+1}}{\alpha_{\ell}}
\end{picture} & \mbox{if 
$\alpha_i < \beta,\; \gamma \leq \alpha_{i+1}$,} \\
\setlength{\unitlength}{5mm}
\begin{picture}(15,2)(0,0.8)
\dominocrossbarx{\alpha_i}{\alpha_\ell}
{\beta}{\alpha_{1}}{\alpha_{i-1}}{\gamma}{\alpha_{i+1}}{\alpha_{\ell -1}}
\end{picture} & \mbox{if 
$
\beta \leq \alpha_1, \, 
\alpha_i < \gamma \leq \alpha_{i+1},
$} \\
\setlength{\unitlength}{5mm}
\begin{picture}(15,2)(0,0.8)
\dominocrossbar{\alpha_\ell}{\gamma}{\beta}{\alpha_{1}}{\alpha_{\ell -1}}
\end{picture} & \mbox{if 
$
\beta \leq \alpha_1, \, \gamma > \alpha_\ell .
$} \\
\end{cases}
\end{align*}
The first and the second ones are associated with
$B_{(\ell,1,1)}$, and the third and fourth ones 
with $B_{(\ell+1,1)}$.
We shall omit explicit expressions for 
$\iota^{-1},(\iota')^{-1}$ but give
one for $(\iota'')^{-1}$ in section \ref{sec:6}.

Now we consider a tensor product of $\mathcal{N}$ crystals
$\mathcal{B}:=B_{\lambda_1} \otimes \cd \otimes 
B_{\lambda_\mathcal{N}}$ where
each $\lambda_i$ ($1 \leq i \leq \mathcal{N}$) is 
a rectangle.
{We let $\sigma$ denote the map for the
isomorphism between tensor products
of any two such crystals of rectangular shape
Young tableaux.}
\begin{definition}\label{def:nov2_1}
For $\bp = b_1 \ot \cd \ot b_\mathcal{N} \in \mathcal{B}$
we define $\sigma_i$ ($1 \leq i \leq \mathcal{N}-1$) by
\begin{displaymath}
\sigma_i(\bp) = b_1 \ot \cd \ot 
\sigma(b_{i} \ot b_{i+1}) \ot
\cd \ot b_\mathcal{N}.
\end{displaymath}
\end{definition}
Then we have
\begin{proposition}\label{prop:oct9_10}
{
Fix a positive integer $\ell$ and
consider a tensor product
$\mathcal{B}=B_{\lambda_1} \otimes \cd \otimes 
B_{\lambda_\mathcal{N}}$ with
each $\lambda_i$ ($1 \leq i \leq \mathcal{N}$) being one of
$(\ell), (1), (1,1)$.}
Then the $\sigma_i$'s generate the symmetric group, i.\ e.\
\begin{align*}
& \sigma_i^2 = {\rm Id}, \\
&\sigma_i \sigma_j = \sigma_j \sigma_i \quad \mbox{for} \quad 
|i-j| \geq 2,\\
&\sigma_i \sigma_{i+1} \sigma_i =\sigma_{i+1} \sigma_i \sigma_{i+1}.
\end{align*}
\end{proposition}
\begin{proof}
Since $\sigma_i$ is the unique isomorphism the first identity holds.
The second identity is obvious.
We prove the third (Yang-Baxter) identity.
Consider {any such}
$\mathcal{B}=B_{\lambda_1} \otimes B_{\lambda_2} \otimes B_{\lambda_3}$
{where} 
$\lambda_1, \lambda_2, \lambda_3$ are all different.
It is sufficient to check the {identity} on {this} 
$\mathcal{B}$.
Moreover we can set $\lambda_1=(\ell), \lambda_2=(1)$ and $\lambda_3=(1,1)$
since the other cases are derived from it, by applying suitable
compositions of $\sigma_1$ and
$\sigma_2$.
Let $b_1 \otimes b_2 \otimes b_3$ be any element of $\mathcal{B}$.
There are
an $sl_n$ highest weight element $u$ in $\mathcal{B}$ and
a sequence $i_1, \ldots, i_m $ for some $m \in \Zn$
such that $b_1 \otimes b_2 \otimes b_3 = \ft{i_m} \cd \ft{i_1} u$.
Therefore it is sufficient to prove 
\begin{equation}\label{eq:oct9_1}
\sigma_{2} \sigma_1 \sigma_{2} \sigma_1 \sigma_{2} \sigma_1 u = u
\end{equation}
for every $sl_n$ highest weight element in $\mathcal{B}$,
since then we have
\begin{math}
\sigma_{2} \sigma_1 \sigma_{2} \sigma_1 \sigma_{2} \sigma_1 
(b_1 \otimes b_2 \otimes b_3) = 
\sigma_{2} \sigma_1 \sigma_{2} \sigma_1 \sigma_{2} \sigma_1
\ft{i_m} \cd \ft{i_1} u 
= \ft{i_m} \cd \ft{i_1} 
\sigma_{2} \sigma_1 \sigma_{2} \sigma_1 \sigma_{2} \sigma_1
u 
= \ft{i_m} \cd \ft{i_1} u =b_1 \otimes b_2 \otimes b_3.
\end{math}
By the LR rule \cite{Fl,Sa} we have
\begin{displaymath}
B_{(\ell)} \ot B_{(1)} \ot B_{(1,1)} = 
B_{(\ell,1,1,1)} \oplus B_{(\ell, 2, 1)} \oplus
B_{(\ell+1, 2)} \oplus B_{(\ell+2, 1)} \oplus 
(B_{(\ell+1,1,1)})^{\oplus 2}
\end{displaymath}
Thus for $\lambda = (\ell,1,1,1),(\ell, 2, 1),(\ell+1, 2)$
or $(\ell+2, 1)$ its associated
$sl_n$ highest weight element is unique and hence
(\ref{eq:oct9_1}) follows.
For $\lambda = (\ell+1,1,1)$ there are two highest weight elements,
for which we can verify (\ref{eq:oct9_1}) directly
(See Appendix \ref{app:b}).
\end{proof}

\begin{remark}\label{rem:oct24_2}
According to \cite{S,SW} the Yang-Baxter identity holds for
any three $sl_n$ crystals of rectangular shape Young tableaux,
{hence the above Proposition holds
with each $\lambda_i$ being a generic rectangle}.
In the above 
we gave an elementary proof of this identity in the special case
needed for us.
\end{remark}
\section{Main theorem for the basic case}\label{sec:5}
{The highest weight elements of $B_\ell$ and $B_\natural$ are
given by}
\begin{align*}
u_\ell &=
\setlength{\unitlength}{5mm}
\begin{picture}(6,1.5)(0.5,0.3)
\put(1,0){\line(1,0){5}}
\put(1,1){\line(1,0){5}}
\put(1,0){\line(0,1){1}}
\put(2,0){\line(0,1){1}}
\put(3,0){\line(0,1){1}}
\put(5,0){\line(0,1){1}}
\put(6,0){\line(0,1){1}}
\put(1,0){\makebox(1,1){$1$}}
\put(2,0){\makebox(1,1){$1$}}
\put(3,0){\makebox(2,1){$\cd$}}
\put(5,0){\makebox(1,1){$1$}}
\end{picture}
\in B_\ell,
\\ 
u_\natural &=
\setlength{\unitlength}{5mm}
\begin{picture}(3,2)(-0.7,0.7)
\multiput(0,0)(1,0){2}{\line(0,1){2}}
\multiput(0,0)(0,1){3}{\line(1,0){1}}
\put(0,0){\makebox(1,1){${2}$}}
\put(0,1){\makebox(1,1){${1}$}}
\end{picture}
\in B_\natural.
\end{align*}
Define the set of basic paths by
\begin{displaymath}
\P = \left\{ \bp = p_1 \ot p_2 \ot \cd \in B_1^{\otimes \infty}
| p_i = \singlebox{1} \, \mbox{ for $i \gg 1$} \right\}.
\end{displaymath}
{A basic path is regarded as an 
infinite array of boxes of capacity one
with finite number of balls scattered
among them, where $\singlebox{1}$ represents
an empty box and $\singlebox{\alpha} \; (\alpha \geq 2)$
a box containing a ball with index $\alpha$.}
{We adopted a set of half-infinite paths as 
the space for the states of the automaton, 
but the formulation here is
essentially not different from that in section \ref{sec:2}.}

Let $B'_\natural$ be a subset of $B_\natural$ whose
every element has a ``1".
We introduce operators $T_\ell \; (\ell \geq 1)$ 
and $T_\natural$ on $\P$ as follows.
For any $\bp \in \P$ we define $T_\ell (\bp) \in \P$ 
and $T_\natural (\bp) \in \P$ by
using the maps for the $sl_n$ crystal isomorphism as
\begin{align}
u_\ell \ot \bp & \stackrel{\sim}{\mapsto} T_\ell (\bp) \ot u_\ell,\nonumber\\
u_\natural \ot \bp & \stackrel{\sim}{\mapsto} T_\natural (\bp) \ot b(\bp).
\label{eq:oct13_3}
\end{align}
Here $b(\bp) \in B'_\natural$ 
depends on the path $\bp$.
The application of $T_\natural$ on a path
is equivalent to the procedure conducted by the decoding carrier in 
section \ref{sec:2}, and
we can regard $b(\bp)$ as 
{an outgoing carrier} which {\em takes off}
a letter $\geq 2$. (See Example \ref{ex:oct10_12}.)
{In what follows we occasionally use
this terminology in some cases, and in particular we say {that}
{\em the decoding carrier takes off a ``2"
(resp.~a letter $\geq 3$)}
when $b(\bp)=u_\natural$ (resp.~$b(\bp) \ne u_\natural$).}

The operator $T_\ell$ gives a time evolution of the
automaton that can be described by a carrier 
of capacity $\ell$ \cite{TM,FOY}.
For $\ell = \infty$ we have
\begin{proposition}[\cite{FOY}]\label{pr:oct29_1}
The operator $T_\infty$ gives the {same}
time evolution of the basic coloured system 
{as} in Definition
\ref{def:oct22_1}.
\end{proposition}
For $\bp= p_1 \ot p_2 \ot \cd  \in \P$ let $F = F(\bp)=
\max\{ i | p_i \ne \singlebox{1} \}$ be the position of the
rightmost non-empty box.
{We shall describe the map $\iota'$ in section \ref{sec:4}
in terms of the processes depicted 
in (\ref{eq:oct10_11}),(\ref{eq:oct10_10}).}
It is easy to see that
\begin{lemma}\label{lem:nov1_1}
When $T_\natural$ is applied on $\bp$,

\begin{enumerate}
\item
The possible process
that occurs at the position $F+1$ is $a$ or $b$ in 
(\ref{eq:oct10_11}).
\item 
The possible process
that occurs at every position $\geq F+2$ is $a$ in 
(\ref{eq:oct10_11}).
\end{enumerate}
\end{lemma}
Then we have
\begin{lemma}\label{lem:nov1_2}
Apply $T_\natural$ on $\bp$.
If the decoding carrier takes off a ``2" 
then $F(T_\natural(\bp)) = F(\bp)$.
\end{lemma}
\begin{proof}
{}From the above Lemma the only possible
process that occurs at $F+1$ is $a$ in 
(\ref{eq:oct10_11}) in this case.
\end{proof}
\begin{lemma}\label{lem:oct23_2}
Suppose all the decoding carriers take off ``2"s
when $(T_\natural)^k$ is applied on $\bp$.
Then there is no letter $\geq 3$ 
at the positions $\geq F-k+1$ in $\bp$.
\end{lemma}
\begin{proof}
Induction on $k$.
Suppose there is a letter $\geq 3$, {say $\gamma$}, 
at the position $F$.
By applying $T_\natural$ on $\bp$,
the possible process that occurs at $F$ is one of $c$ - $g$
in (\ref{eq:oct10_11}), (\ref{eq:oct10_10}). 
Hence the $\gamma$ is loaded into the carrier in any case.
Then by Lemma \ref{lem:nov1_1}
the decoding carrier should 
take off a letter $\geq 3$.
(See the following Example.)

Suppose there is a letter $\geq 3$, say $\gamma$, 
at the position $F-k+1$ and
the decoding carrier takes off a ``2"
when $T_\natural$ is applied on $\bp$.
If so, this $\gamma$ should go rightwards
by at least one box position {at this time.
Therefore there is at least one letter $\geq 3$ at 
some position $\geq F-k+2$ in $T_\natural(\bp)$.}
{The proof follows by}
induction and Lemma \ref{lem:nov1_2}.
\end{proof}
\begin{example}
{}From (\ref{eq:oct10_11}) and (\ref{eq:oct10_10}) 
the possible loading-unloading processes which can occur at
$F$ and $F+1$ are as follows.
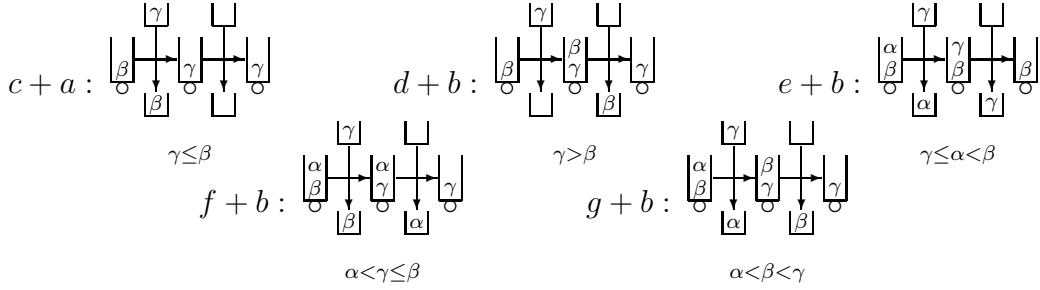
\begin{displaymath}
c+a:\;
\carrierpicturedouble{}{\beta}{\gamma}
{\beta}{}{\gamma}{}{\gamma}{\gamma \leq \beta}\qquad \qquad
d+b:\;
\carrierpicturedouble{}{\beta}{\gamma}
{}{\beta}{\gamma}{\beta}{\gamma}{\gamma > \beta}\qquad \qquad
e+b:\;
\carrierpicturedouble{\alpha}{\beta}{\gamma}
{\alpha}{\gamma}{\beta}{\gamma}{\beta}{\gamma \leq \alpha < \beta}
\end{displaymath}
\begin{displaymath}
f+b:\;
\carrierpicturedouble{\alpha}{\beta}{\gamma}
{\beta}{\alpha}{\gamma}{\alpha}{\gamma}{\alpha < \gamma \leq \beta}
\qquad\qquad
g+b:\;
\carrierpicturedouble{\alpha}{\beta}{\gamma}
{\alpha}{\beta}{\gamma}{\beta}{\gamma}{\alpha < \beta < \gamma}
\end{displaymath}
\vspace{5mm}
\par\noindent
Therefore if $\gamma \geq 3$ then the {decoding} 
carrier takes off a letter $\geq 3$.
\end{example}
\vspace{5mm}
\par\noindent
By taking $k=F$ we have
\begin{coro}\label{lem:oct23_1}
Suppose all the decoding carriers take off ``2"s
when $(T_\natural)^F$ is applied on $\bp$.
Then there is no letter $\geq 3$ in $\bp$.
\end{coro}
{}From {this Corollary} we can deduce that
\begin{theorem}\label{th:oct9_1}
For any $\bp \in \P$ all the letters $\geq 3$ 
in $\bp$ can be removed
by applying $T_\natural$ sufficiently many times.
\end{theorem}
%
Thus for any $\bp \in \P$ 
{we can find a non-negative integer $N$} 
such that both $\tilde{\bp}:=(T_\natural)^N (\bp)$ and
$\widetilde{T_\ell(\bp)}:=(T_\natural)^N (T_\ell(\bp))$ 
are composed of only $\singlebox{1}$ and $\singlebox{2}$.
Choose any integer $N$ {such that
these conditions are satisfied}.
{By using
(\ref{eq:oct13_3}) we have}
\begin{align}\label{eq:oct13_1}
u_\natural^{\ot N} \ot \bp & \stackrel{\sim}{\mapsto} \tilde{\bp} \ot
\left( b_1 \ot \cdots \ot b_N \right), \\
u_\natural^{\ot N} \ot T_\ell(\bp) & \stackrel{\sim}{\mapsto} 
\widetilde{T_\ell(\bp)} \ot
\left( b'_1 \ot \cdots \ot b'_N \right), \nonumber
\end{align}
where $b_i,b'_i \in B'_\natural \; (1 \leq i \leq N)$ 
are defined as
$b_i = b((T_\natural)^{N-i} (\bp))$ and 
$b'_i = b((T_\natural)^{N-i}(T_\ell (\bp)))$.
It is easy to see that
if $\bp \in \P$ has no letter $\geq 3$ 
then $T_\natural(\bp)=\bp$.
Thus the $\tilde{\bp}$
does not depend on the choice of $N$.

Now we present the main theorem in this paper.
\begin{theorem}\label{th:oct13_2}
With the notation as above
the following relations hold:
$T_\ell(\tilde{\bp}) = \widetilde{T_\ell(\bp)}$ and
$b_i = b'_i$ for $1 \leq i \leq N$.
\end{theorem}
\begin{proof}
By the isomorphism of $sl_n$ crystal we have
\begin{align*}
u_\ell \ot u_\natural^{\ot N} \ot \bp & \stackrel{\sim}{\mapsto}
u_\natural^{\ot N} \ot u_\ell \ot \bp \\
& \stackrel{\sim}{\mapsto} 
u_\natural^{\ot N}  \ot T_\ell(\bp) \ot u_\ell\\
& \stackrel{\sim}{\mapsto} 
\widetilde{T_\ell(\bp)} \ot
\left( b'_1 \ot \cdots \ot b'_N \right) \ot u_\ell,
\end{align*}
and
\begin{align*}
u_\ell \ot u_\natural^{\ot N} \ot \bp & \stackrel{\sim}{\mapsto}
u_\ell \ot \tilde{\bp} \ot \left( b_1 \ot \cdots \ot b_N \right)\\
& \stackrel{\sim}{\mapsto}
T_\ell(\tilde{\bp}) \ot u_\ell \ot \left( b_1 \ot \cdots \ot b_N \right)\\
& \stackrel{\sim}{\mapsto}
T_\ell(\tilde{\bp}) \ot \left( b_1 \ot \cdots \ot b_N \right)\ot u_\ell .
\end{align*}
{For any positive integer $L$
we consider a tensor product of crystals} 
$\mathcal{B} = B_\ell \ot B_\natural^{\ot N} \ot B_1^{\ot L} \simeq 
B_1^{\ot L} \ot B_\natural^{\ot N} \ot B_\ell $.
{For this $\mathcal{B}$ define
$\sigma_i$ as in Definition \ref{def:nov2_1}
with $\mathcal{N}=L+N+1$.}
To prove the Theorem it is sufficient to show
the following $x$ and $y$ coincide each other.
\begin{align*}
x:&=
(\sigma_L \cd \sigma_2 \sigma_1) (\sigma_{L+1} \cd \sigma_3 \sigma_2) \cd
(\sigma_{N+L-1} \cd \sigma_{N+1} 
\sigma_N ) (\sigma_{N+L} \cd \sigma_2 \sigma_1),\\
y:&=
(\sigma_{N+L} \cd \sigma_2 \sigma_1)
(\sigma_{L+1} \cd \sigma_3 \sigma_2)(\sigma_{L+2} \cd \sigma_4 \sigma_3) \cd
(\sigma_{N+L} \cd \sigma_{N+2} \sigma_{N+1} ).
\end{align*}
They serve as two possible ways to send an element of 
$B_\ell \ot B_\natural^{\ot N} \ot B_1^{\ot L} $
into 
$B_1^{\ot L} \ot B_\natural^{\ot N} \ot B_\ell $.
The $x$ is 
for
$B_\ell \ot B_\natural^{\ot N} \ot B_1^{\ot L} 
\stackrel{\sim}{\rightarrow}  
B_\natural^{\ot N} \ot B_\ell \ot B_1^{\ot L} 
\stackrel{\sim}{\rightarrow} 
B_\natural^{\ot N} \ot B_1^{\ot L} \ot B_\ell 
\stackrel{\sim}{\rightarrow} 
B_1^{\ot L} \ot B_\natural^{\ot N} \ot B_\ell$
and the $y$ is for
$B_\ell \ot B_\natural^{\ot N} \ot B_1^{\ot L} 
\stackrel{\sim}{\rightarrow}  
B_\ell \ot B_1^{\ot L} \ot B_\natural^{\ot N} 
\stackrel{\sim}{\rightarrow}  
B_1^{\ot L} \ot B_\ell \ot B_\natural^{\ot N} 
\stackrel{\sim}{\rightarrow}  
B_1^{\ot L} \ot B_\natural^{\ot N} \ot B_\ell$.
By Proposition \ref{prop:oct9_10}
one can verify $x=y$ directly.
(See the following Example.)
\end{proof}
\begin{example}
Let $N=L=3$.
For notational simplicity we denote $\sigma_i$ by $i$.
By repeated use of
$\sigma_i \sigma_{i+1} \sigma_i =\sigma_{i+1} \sigma_i \sigma_{i+1}$
we have $1234321 = 4321234$ for instance.
Then
\begin{align*}
(32{\bf 1})(43{\bf 2})(54{\bf 3})(65{\bf 4321}) &= 
(3{\bf 2})(4{\bf 3})(5{\bf 4})(6{\bf 5432}1)(234) \\
&=({\bf 3})({\bf 4})({\bf 5})({\bf 6543}21)(345)(234) \\
&=(654321)(456)(345)(234) \\
&=(654321)(432)(543)(654).
\end{align*}
Here we used $\sigma_i \sigma_j = \sigma_j \sigma_i$ for $|i-j|\geq 2$.
\end{example}
Let $y_i(\ne 1)$ be the letter that appears in
the $b_i \; (1 \leq i \leq N)$ in (\ref{eq:oct13_1}).
Denote the word $y_1 \ldots y_N$ by $\by$ and write
the relation (\ref{eq:oct13_1}) symbolically as 
$\bp = \tilde{\bp}\oplus \by$.
Then by Theorem \ref{th:oct13_2} we have
\begin{equation} 
T_\ell(\bp) = T_\ell(\tilde{\bp}) \oplus \by,
\end{equation}
for any $\ell$.
Thus by letting $\ell$ be infinity we obtain (\ref{eq:oct10_1})
by Proposition \ref{pr:oct29_1}.
\section{Generalization to the inhomogeneous case}\label{sec:6}
For the inhomogeneous case we need to prove the
Yang-Baxter identity on $B_{\ell_1} \otimes B_{\ell_2} \otimes B_{\ell_3}$
and on $B_{\ell_1} \otimes B_{\ell_2} \otimes B_\natural$.
{Although there is a more general assertion mentioned in
Remark \ref{rem:oct24_2},}
we shall still give some
explanations on this identity in these special cases
in order to make our arguments self-contained.

As a set the $sl_n$ crystal 
$B_\ell$ in (\ref{eq:oct24_1}) is also given by
\begin{equation}\label{eq:oct24_3}
B_\ell = \left\{ (x_1,\cdots,x_n) \in
\Zn^{n} \bigg| \sum_{i=1}^n x_i= \ell 
\right\}.
\end{equation}
Here $x_i$ shows the number of the letter $i$ that appears in
the tableau representation (\ref{eq:oct24_1}).
Let $B$ and $B'$ be any two of these $sl_n$ crystals.
Given $x=(x_1,\ldots,x_n) \in B, y=(y_1,\ldots,y_n) \in B'$
we define a piecewise linear map
$R:(x,y) \mapsto (x',y')$ as
$x'=(x'_1,\ldots,x'_n), y'=(y'_1,\ldots,y'_n)$ and
\begin{math}
x_i' = y_i + P_{i+1} - P_i,\;
y_i' = x_i + P_i - P_{i+1},
\end{math}
where
\begin{math}
P_i = 
\max_{1 \leq j \leq n}
\left( \sum_{k=1}^{j-1} (y_{k+i-1} - x_{k+i-1}) + 
y_{j+i-1} \right).
\end{math}
Here the indices are interpreted in modulo $n$.
This piecewise linear map gives
the isomorphism of $sl_n$ crystals 
$B \ot B' \stackrel{\sim}{\rightarrow} B' \ot B$ \cite{HHIKTT}.
Consider such
$\mathcal{B}=B_{\ell_1} \otimes B_{\ell_2} \otimes B_{\ell_3}$.
{For this $\mathcal{B}$ define
$\sigma_i$ as in Definition \ref{def:nov2_1}
with $\mathcal{N}=3$ and $\sigma = R$.
Then the Yang-Baxter identity 
$\sigma_1 \sigma_{2} \sigma_1=\sigma_{2} \sigma_1 \sigma_{2}$ 
holds on this $\mathcal{B}$.
A simple proof of this identity on this $\mathcal{B}$ 
is given
by using a birational map analogue of the above $R$
\cite{KNY,Y}.}

Next we consider such
$\mathcal{B}=B_{\ell_1} \otimes B_{\ell_2} \otimes B_{\natural}$.
Define $\sigma_i$ as above but now 
$\sigma$ is one of
$R$, $\iota'', (\iota'')^{-1}$.
\begin{lemma}\label{lem:oct24_4}
The identity 
$\sigma_1 \sigma_{2} \sigma_1=\sigma_{2} \sigma_1 \sigma_{2}$ 
holds on $\mathcal{B}=B_{\ell_1} \otimes 
B_{\ell_2} \otimes B_{\natural}$.
\end{lemma}
\begin{proof}
For simplicity we assume $\ell_1 > \ell_2$.
It is sufficient to prove 
\begin{math}
\sigma_{2} \sigma_1 \sigma_{2} \sigma_1 \sigma_{2} \sigma_1 u = u
\end{math}
for every $sl_n$ highest weight element $u$ in $\mathcal{B}$.
By the LR rule \cite{Fl,Sa} we have
\begin{align*}
B_{(\ell_1)} \ot B_{(\ell_2)} \ot B_{(1,1)} &= 
\left( \bigoplus_{x=1}^{\ell_2} B_{(\ell_1+\ell_2-x,x,1,1)}
\right) \oplus
\left( \bigoplus_{x=0}^{\ell_2} B_{(\ell_1+\ell_2-x+1,x+1)}
\right) \\
& \oplus 
\left( \bigoplus_{x=0}^{\ell_2-1} (B_{(\ell_1+\ell_2-x,x+1,1)})^{\oplus 2}
\right) \oplus
B_{(\ell_1,\ell_2+1,1)}.
\end{align*}
So it is enough to prove the above identity only for
the highest weight elements associated with $B_{(\ell_1+\ell_2-x,x+1,1)}$
for $0 \leq x \leq \ell_2-1$ because the other 
$B_\lambda$s are multiplicity free.
It can be verified directly.
(See Appendix \ref{app:c}.)
\end{proof}
For a given sequence of positive integers 
$\boldsymbol{\ell} = (\ell_1, \ell_2, \cd)$
define the associated
set of inhomogeneous paths by
\begin{displaymath}
\P_{\boldsymbol{\ell}} 
= \left\{ \bp = p_1 \ot p_2 \ot \cd \in B_{\ell_1} \ot
B_{\ell_2} \ot \cd | p_i = u_{\ell_i} \, \mbox{ for $i \gg 1$} \right\}.
\end{displaymath}
An inhomogeneous 
path is regarded as an infinite array of boxes of various capacities
with finite number of balls scattered
among them.
At the $i$th position there is a box $p_i$ of capacity $\ell_i$.
If it has the expression $p_i=(x_1,\cd,x_n)$ then {we interpret
it as} a box containing $x_k$ balls with index $k$ for $2 \leq k \leq n$.
We define the operators $T_\ell \; (\ell \geq 1)$ 
and $T_\natural$ which act on $\P_{\boldsymbol{\ell}}$ 
by using the same formulas in (\ref{eq:oct13_3}).
As in the basic case the operator $T_\ell$ gives the time evolution
described by a carrier of capacity $\ell$ \cite{HHIKTT}.
For $\ell = \infty$ the operator $T_\infty$ also has a description
that is a generalization of
Definition \ref{def:oct22_1} to the inhomogeneous
case \cite{TTM}.

We gave an explicit formula for the
map $\iota'': B_\ell \ot B_\natural \stackrel{\sim}{\rightarrow} 
B_\natural \ot B_\ell$ in section \ref{sec:4}.
The inverse of $\iota''$ is given as follows
\begin{align*}
(\iota'')&^{-1}
:\domino{\alpha}{\beta} \;\ot\;
\threelettersinbarx{\gamma_1}{\gamma_2}{\gamma_\ell}
\nonumber\\
\stackrel{\sim}{\mapsto}&
\begin{cases}
\threelettersinbar{\gamma_2}{\gamma_\ell}{\alpha}
\ot
\domino{\gamma_1}{\beta} \;
\mbox{if} \, \gamma_\ell \leq \alpha, & {\rm (I)}\\
\sixlettersinbarx{\gamma_2}{\gamma_{i-1}}{\alpha}{\gamma_{i+1}}
{\gamma_{\ell}}{\beta}
\ot
\domino{\gamma_1}{\gamma_i} \;
\mbox{if} \, \gamma_{i-1} \leq \alpha < \gamma_i,
\gamma_\ell \leq \beta, & {\rm (II)} \\
\threelettersinbar{\gamma_2}{\gamma_\ell}{\beta}
\ot
\domino{\alpha}{\gamma_1} \;
\mbox{if} \, \alpha < \gamma_1,
\gamma_\ell \leq \beta, & {\rm (III)} \\
\sixlettersinbar{\gamma_{i-1}}{\alpha}{\gamma_{i+1}}
{\gamma_{j-1}}{\beta}{\gamma_{j+1}} 
\ot
\domino{\gamma_i}{\gamma_j} \;
\mbox{if} \, \gamma_{i-1} \leq \alpha < \gamma_i,
\gamma_{j-1} \leq \beta < \gamma_j, & {\rm (IV)} \\
\fivelettersinbar{\gamma_1}{\gamma_{i-1}}{\alpha}{\gamma_{i+1}}
{\gamma_{\ell}} 
\ot
\domino{\beta}{\gamma_i} \;
\mbox{if} \, \gamma_{i-1} \leq \alpha, \beta < \gamma_i.
& {\rm (V)}
\end{cases}
\end{align*}
%
This formula shows the
loading-unloading processes of the decoding carrier
at a box with capacity $\geq 2$,
corresponding to those pictures in
(\ref{eq:oct10_11}), (\ref{eq:oct10_10}) 
for the case of capacity one.
As in the basic case let $F$
be the position of the rightmost non-empty box 
i.\ e.\ $F = F(\bp)=\max\{ i | p_i \ne u_{\ell_i} \}$ for
$\bp= p_1 \ot p_2 \ot \cd  \in \P_{\boldsymbol{\ell}}$.
It is easy to see that
\begin{lemma}\label{lem:nov2_1}
When $T_\natural$ is applied on $\bp$,

\begin{enumerate}
\item
The possible process
that occurs at the position $F+1$ is (I).
\item 
The possible process
that occurs at every position $\geq F+2$ is (I)
with $\alpha = 1$.
\end{enumerate}
\end{lemma}
Then we have
\begin{lemma}\label{lem:nov2_2}
Apply $T_\natural$ on $\bp$.
If the decoding carrier takes off a ``2"
then $F(T_\natural(\bp)) = F(\bp)$.
\end{lemma}
\begin{proof}
{}From the above Lemma the only possible
process that occurs at $F+1$ is (I) with $\alpha = 1$
in this case.
\end{proof}
\begin{lemma}\label{lem:oct25_6}
{Suppose all 
the decoding carriers take off ``2"s
when $(T_\natural)^{\ell_F + \ell_{F-1} + \cd + \ell_{F-k+1}}$ 
is applied on $\bp$.
Then there is no letter $\geq 3$ 
at the positions $\geq F-k+1$ in $\bp$.}
\end{lemma}
\begin{proof}
Induction on $k$.
Suppose all 
the decoding carriers take off ``2"s
when $(T_\natural)^{\ell_F}$ 
is applied on $\bp$.
If so, no carrier that leaves from the position $F$ 
has a letter $\geq 3$, by Lemma \ref{lem:nov2_1}.
This implies that under the application of $(T_\natural)^{\ell_F}$
the possible processes that occur at the position $F$ 
are (I), (II), (III), and not (IV) or (V).
Now suppose there is a letter $\geq 3$, say $\gamma$,
in the tableau at the position $F$.
In (I), (II), (III)
any letter in the tableau
is either loaded into the carrier or 
shifted to the left inside the tableau.
In particular the leftmost letter $\gamma_1$ is
always loaded into the carrier.
Therefore the $\gamma$
should be loaded into some carrier
when $(T_\natural)^{\ell_F}$ is applied on $\bp$,
because the width of the tableau is $\ell_F$.
This contradicts to the above observation.
Thus there is no letter $\geq 3$ at the position $F$.

{Suppose there is a letter $\geq 3$, say $\gamma$, at 
the position $F-k+1$ and 
{all the decoding carriers take off ``2"s}
when $(T_\natural)^{\ell_{F-k+1}}$ is
applied on $\bp$.
Suppose also that the processes that occur at the position $F-k+1$
are (I),(II),(III), and not (IV) or (V) at this time.
If so, the $\gamma$ 
should be loaded into some carrier
when $(T_\natural)^{\ell_{F-k+1}}$ is applied on $\bp$,
by the same reason in the previous paragraph.
If not, there should be at least one letter $\geq 3$ delivered to some
position $\geq F-k+2$ by a carrier which
has left from the position $F-k+1$ after the process (IV) or (V).
In any case 
there should be at least one letter $\geq 3$ 
at some position $\geq F-k+2$ in $(T_\natural)^{\ell_{F-k+1}}(\bp)$.
The proof follows by
induction and Lemma \ref{lem:nov2_2}.}
\end{proof}
By taking $k=F$ we have
\begin{coro}
{Suppose all 
the decoding carriers take off ``2"s
when $(T_\natural)^{\ell_F + \ell_{F-1} + \cd + \ell_{1}}$ 
is applied on $\bp$.
Then there is no letter $\geq 3$ in $\bp$.}
\end{coro}
Therefore
\begin{lemma}
For any 
$\bp \in \P_{\boldsymbol{\ell}}$ all the letters $\geq 3$ in $\bp$
can be removed
by applying $T_\natural$ sufficiently many times.
\end{lemma}
By this Lemma and the Yang-Baxter identity we {can deduce} that
the separation of colour degree of freedom
(Theorem \ref{th:oct13_2})
also holds in the inhomogeneous case.

\vspace{0.4cm}
\noindent
\textbf{Acknowledgements} \hspace{0.1cm}
The author thanks Atsuo Kuniba, Masato Okado,
and Yasuhiko Yamada for valuable discussions.
\clearpage
\appendix
\section{An example of 
$T(\bp) = T(\tilde{\bp}) \oplus \by$}\label{app:a}
For $t=1$ of Example \ref{ex:oct3_2}
we obtain the following separation.
\begin{verbatim}
       s=0   .....55432...542..2.............. 2
       s=1   ......55422...532.4.............. 4
       s=2   .......55222...4325.............. 5
       s=3   ........52222..543.2............. 2
       s=4   .........22222.554.3............. 3
       s=5   .........22222..5524............. 4
       s=6   .........22222...5252............ 5
       s=7   .........22222....2522........... 5
       s=8   .........22222....2.222.......... 
\end{verbatim}
For $t=2$ we obtain the following.
\begin{verbatim}
       s=0   ..........55432.54.22............ 2
       s=1   ...........55422.5342............ 4
       s=2   ............55222.4532........... 5
       s=3   .............522225.432.......... 2
       s=4   ..............222252543.......... 3
       s=5   ..............2222.25542......... 4
       s=6   ..............2222.2.5522........ 5
       s=7   ..............2222.2..5222....... 5
       s=8   ..............2222.2...2222...... 
\end{verbatim}
For $t=3$ we obtain the following.
\begin{verbatim}
       s=0   ...............5435..54222....... 2
       s=1   ................5423.55422....... 4
       s=2   .................5242.55322...... 5
       s=3   ..................2522.54322..... 2
       s=4   ..................2.22255432..... 3
       s=5   ..................2.222.55422.... 4
       s=6   ..................2.222..55222... 5
       s=7   ..................2.222...52222.. 5
       s=8   ..................2.222....22222. 
\end{verbatim}
Note that we always obtain the same sequence of removed letters
$\by = 55432542$ (in reverse order), and
that the automaton states in the
last rows coincide with the states in Example \ref{ex:oct3_1}.
\section{Proof of Proposition \ref{prop:oct9_10}
(Continued)}\label{app:b}

For each highest element $u$ of the shape $\lambda = (\ell+1,1,1)$ 
the relation
$\sigma_{2} \sigma_1 \sigma_{2} \sigma_1 \sigma_{2} \sigma_1 u = u
$
can be verified as
(we have set $\ell = 3$ here)
\begin{align*}
& \threeboxes{1}{1}{1} \ot \singlebox{1} \ot \domino{2}{3} \\
& \stackrel{\sim}{\mapsto}
\singlebox{1} \ot \threeboxes{1}{1}{1} \ot \domino{2}{3} \stackrel{\sim}{\mapsto}
\singlebox{1} \ot \domino{1}{2} \ot \threeboxes{1}{1}{3} \stackrel{\sim}{\mapsto}
\domino{1}{2} \ot \singlebox{1} \ot \threeboxes{1}{1}{3} \\
&\stackrel{\sim}{\mapsto}
\domino{1}{2}\ot \threeboxes{1}{1}{1}  \ot \singlebox{3} \stackrel{\sim}{\mapsto}
\threeboxes{1}{1}{1}  \ot \domino{1}{2}\ot \singlebox{3} \stackrel{\sim}{\mapsto}
\threeboxes{1}{1}{1} \ot \singlebox{1} \ot \domino{2}{3},
\end{align*}
\begin{align*}
& \threeboxes{1}{1}{1} \ot \singlebox{2} \ot \domino{1}{3} \\
& \stackrel{\sim}{\mapsto}
\singlebox{1} \ot \threeboxes{1}{1}{2} \ot \domino{1}{3} \stackrel{\sim}{\mapsto}
\singlebox{1} \ot \domino{2}{3} \ot \threeboxes{1}{1}{1} \stackrel{\sim}{\mapsto}
\domino{1}{2} \ot \singlebox{3} \ot \threeboxes{1}{1}{1} \\
&\stackrel{\sim}{\mapsto}
\domino{1}{2}\ot \threeboxes{1}{1}{3}  \ot \singlebox{1} \stackrel{\sim}{\mapsto}
\threeboxes{1}{1}{1}  \ot \domino{2}{3}\ot \singlebox{1} \stackrel{\sim}{\mapsto}
\threeboxes{1}{1}{1} \ot \singlebox{2} \ot \domino{1}{3}.
\end{align*}
\section{Proof of Lemma \ref{lem:oct24_4}
(Continued)}\label{app:c}
Recall the description of
the elements of $B_\ell$ in (\ref{eq:oct24_3}).
Denote $(x_1,x_2,x_3,0,\cd,0)$ by $[x_1,x_2,x_3]$.
For each highest element $u$ of the shape $\lambda = (\ell_1+
\ell_2-x,x+1,1)$ 
the relation
$\sigma_{2} \sigma_1 \sigma_{2} \sigma_1 \sigma_{2} \sigma_1 u = u
$
can be verified as
\begin{align*}
& [\ell_1,0,0] \ot [\ell_2-x,x,0] \ot \domino{2}{3} \\
& \stackrel{\sim}{\mapsto}
[\ell_2,0,0] \ot [\ell_1-x,x,0] \ot \domino{2}{3} \stackrel{\sim}{\mapsto}
[\ell_2,0,0] \ot \domino{1}{2} \ot [\ell_1-x-1,x,1] \\
& \stackrel{\sim}{\mapsto}
\domino{1}{2} \ot [\ell_2,0,0] \ot [\ell_1-x-1,x,1] \stackrel{\sim}{\mapsto}
\domino{1}{2}\ot [\ell_1,0,0]  \ot [\ell_2-x-1,x,1] \\
& \stackrel{\sim}{\mapsto}
[\ell_1,0,0]  \ot \domino{1}{2}\ot [\ell_2-x-1,x,1] \stackrel{\sim}{\mapsto}
[\ell_1,0,0] \ot [\ell_2-x,x,0] \ot \domino{2}{3},
\end{align*}
\begin{align*}
& [\ell_1,0,0] \ot [\ell_2-x-1,x+1,0] \ot \domino{1}{3} \\
& \stackrel{\sim}{\mapsto}
[\ell_2,0,0] \ot [\ell_1-x-1,x+1,0] \ot \domino{1}{3} \stackrel{\sim}{\mapsto}
[\ell_2,0,0] \ot \domino{2}{3} \ot [\ell_1-x,x,0] \\
& \stackrel{\sim}{\mapsto}
\domino{1}{2} \ot [\ell_2-1,0,1] \ot [\ell_1-x,x,0] \stackrel{\sim}{\mapsto}
\domino{1}{2}\ot [\ell_1-1,0,1]  \ot [\ell_2-x,x,0] \\
& \stackrel{\sim}{\mapsto}
[\ell_1,0,0]  \ot \domino{2}{3}\ot [\ell_2-x,x,0] \stackrel{\sim}{\mapsto}
[\ell_1,0,0] \ot [\ell_2-x-1,x+1,0] \ot \domino{1}{3}.
\end{align*}

\end{document}